\documentstyle[aps,pra,tighten]{revtex}

\begin{document}
\title{Collapses and Revivals of Collective Excitations in trapped Bose
condensates}
\author{R.Graham\cite{permad}, D.F.Walls and M.J.Collett}
\address{Department of Physics, University of Auckland,
Auckland, New Zealand.}
\vspace{0.5cm}

\author{M.Fliesser}
\address{Fachbereich Physik, Universit\"at - GH Essen, D-45117 Essen, Germany}

\author{E.M.Wright}
\address{Optical Sciences Center, University of Arizona,
        Tucson, AZ 85721, USA.}

\maketitle
\begin{abstract}
We present a quantum theory of low-lying excitations in a trapped Bose
condensate
with finite particle numbers.  We find that even at zero temperature
condensate
number fluctuations and/or fluctuations of the excitation frequency
due to quantum uncertainties of the mode occupation lead to
a collapse of the collective modes due to dephasing. Coherent revivals of the
collective excitations are predicted on a much longer timescale. Depletion of
collective modes due to second-harmonic generation is discussed.
\end{abstract}


\section{Introduction}

In order to investigate the properties of the recently created Bose condensates
of alkali atoms in magnetic traps \cite{1,2,3,4} the excitation of low-lying
collective modes by periodic variations of the trap potential has turned out to
be a very effective tool.  The successful implementation of this idea has
produced results for the collective mode spectrum of the trapped condensate
\cite{5,6}
which are in very good agreement with theoretical predictions based on
solutions of the Gross-Pitaevski equation \cite{7,8,9,Singh} i.e. on
classical mean
field theory at zero temperature. In the experiments, in addition to the mode
frequencies, the decay of the collective excitations in real time could be
observed.  At present there exists no theory which describes the damping of
excitations at finite temperature. In this paper we focus solely on the
decay of the collective exciations at very low (effectively zero) temperature.
In the very low
temperature regime it is difficult to imagine any truly dissipative mechanism
for the observed decay, as long as the trapped system can be considered
closed. The usual dissipative mechanisms in a  homogeneous Bose
condensate
\cite{10} depend on the {\em continuous} mode spectrum found there.

It has recently been shown \cite{WriWalGar96} that nondissipative
interactions
give rise to collapses and revivals of the macroscopic wavefunction
for small atomic condensates, analogous to
those predicted \cite{pred} and observed \cite{obs} for a single mode field
and
a two-level atom. Here we extend this idea to show that very similar
mechanisms may lead to the collapse (apparent damping) of collective
excitations
in a finite Bose condensate with a discrete mode spectrum.  For times much
longer than the observation time in present experiments, this collapse is
predicted to be reversed in `revivals' of the collective excitations.

We shall investigate two possible mechanisms for the collapse. The first
mechanism, discussed in Section II, is based on atom-number
uncertainty in the condensate. It therefore has the same physical origin
as the
collapse of the macroscopic wave function \cite{WriWalGar96}.
However, it will turn out that this
mechanism is much less effective for collective modes. The second
mechanism, described in Section III, is based on a potential dependence of
the mode frequency on the mode occupation. Given such a dependence,
quantum uncertainties in the occupation of a mode must lead to its collapse. A
two-mode model with this property has been discussed recently by Kuklov et.
al. \cite {Kuklov}, while Pitaevskii \cite {Pitaevskii} also recently
discussed
this mechanism in general for nonlinear oscillators and applied it to a
special
collective mode in trapped condensates. Here we wish to present a theory of
the nonlinear self-coupling of the collective hydrodynamic modes in trapped
Bose-condensates and examine the conditions under which the nonlinear
coupling can become effective.

We emphasize that while we propose these collapse-mechanisms as possible
explanations
for the damping of excitations at zero or very low temperatures, in current
experiments there appears to be a substantial finite temperature contribution
which is likely to produce the dominant contribution to the currently
observed damping times \cite {JinMatEns97}.

\section{Collapse due to atom-number uncertainty in the condensate}

The mechanisms we
present are intrinsically quantum mechanical and, in principle, occur in
any {\em
finite} Bose condensed system, in which the spontaneous symmetry
breaking
associated with the {\em infinite} system cannot occur \cite{Gri93}.
For this reason the
usual Bogoliubov analysis of the low-lying excitations of an infinitely
extended
Bose gas, which is based on the assumption of a spontaneously broken
gauge
symmetry, is not directly applicable.  It must be modified by eliminating the
underlying assumption of broken gauge symmetry before the collapse
mechanism
proposed here can be described consistently. Here we first present this
modification of the Bogoliubov theory, then examine the collapse and
revival of the collective excitations.

The grand canonical Hamiltonian of the system is
\cite{Hua87}
\begin{equation}
H = \int d^3 r \left[ \frac
{\hbar^2}{2m} \bbox\nabla\hat\psi^{\dag}\bbox\cdot\bbox\nabla\hat\psi
+ \left( V({\bf r}) + \delta V({\bf r},t) - \mu \right)
\hat\psi^{\dag}\hat\psi
+ \frac{U_0}2 \hat\psi^{\dag 2}\hat\psi^2 \right] \:,
\label{hamiltonian}\end{equation}
where $m$ is the atomic mass, $\mu$ the chemical potential,
$V({\bf r})$ is the static trap potential, $\delta V({\bf r},t)$ its
modulation,
and $U_0=4\pi\hbar^2 a /m$ is proportional to the $s$-wave scattering length
$a$.  The
Bogoliubov
aproach to the low-lying excitations is based on decomposing the boson field
annihilation operator $\hat\psi({\bf r})$ (and its adjoint) as
\cite{Bog47,Bel58}
\begin{equation}
\hat\psi ({\bf r}) = \langle\hat\psi({\bf r})\rangle + \delta\hat\psi({\bf
r}) \:,
\label{decompose} \end{equation}
and then approximating (\ref{hamiltonian}) by a quadratic form in the
operators
$\delta \hat\psi ({\bf r})$ and $\delta \hat\psi^{\dag} ({\bf r})$.  This
approach is successful if the expectation value
$\langle\hat\psi({\bf r})\rangle$ on the right-hand side of
(\ref{decompose}) is
nonzero, which is the case in an infinitely extended system if the U(1) gauge
symmetry is spontaneously broken, or in a finite system if the phase of the
condensate is established with respect to a reference, e.g. by a measurement
of
its phase relative to an infinitely-extended Bose-condensed system \cite{12},
which itself has a well-defined phase.  However, in the experiments on
collective excitations in trapped Bose condensates referred to above a phase
preparation of the condensate is not made, so that
$\langle\hat\psi({\bf r})\rangle=0$ on the right-hand side of
(\ref{decompose})
and this decomposition becomes useless.  In fact, the same kind of problem
appears for a laser far above threshold, where Bose condensation of photons
in
the laser mode occurs, but the phase is not fixed and undergoes a diffusion
process \cite{Haken}.  As is done in that case, it is necessary here to replace
(\ref{decompose}) by  \footnote{The decomposition of $\psi$ and
$\psi^{\dag}$
into amplitude and phase is beset by well-known problems, which arise due
to the
existence of the vacuum state satisfying $\psi|0\rangle=0$.  However, since
$N_0= \int d^3r\rho_0({\bf r})\gg1$ the probability amplitude for the
vacuum
state $|0\rangle$ is negligibly small.}
\begin{equation}
\hat\psi ({\bf r})
= \;e^{i\hat\phi({\bf r})}\sqrt{\rho_0({\bf r})+\delta\hat\rho({\bf
r})}+\hat\psi_1({\bf r})\:.
\label{decompose2}
\end{equation}
The decomposition (3) was used by Popov \cite{Popov} for spatially
homogeneous Bose
condensed systems. Here $\rho_0({\bf r})$ is a $c$-number and
$\delta\hat\rho({\bf r})$
and $\hat\phi({\bf r})$ are operators with commutation relations
$[\delta\hat\rho({\bf r}),e^{i\hat\phi({\bf r}')}]
= -e^{i\hat\phi({\bf r})}\delta^{(3)}({\bf r} - {\bf r}')$
and together they provide a quantum mechanical description of the
condensate and its collective excitations at long wavelengths\cite{Popov}.
It is convenient to decompose $\delta\hat\rho({\bf r})$ further into
\begin{equation}
\delta\hat\rho({\bf r}) = \delta\hat\sigma({\bf r}) +\delta \hat n/\Omega\:,
\label {dec} \end{equation}
where $\delta\hat n= \int d^3r\delta\hat\rho({\bf r},t)$ is canonically
conjugate to the spatially averaged phase variable $\hat{\bar\phi} =
\Omega^{-1}\int d^3r\hat\phi({\bf r})$ according to the commutaion relation
$[\delta\hat{n},\hat{\bar\phi}]=i$ and therefore commutes with ${\bf
\nabla}\hat\phi$ and has discrete, not necessarily
integer eigenvalues with unit spacing. It describes number fluctuations in
the condensate. In the following it is useful to introduce the abbreviation
$\hat\rho({\bf r})=\rho_0({\bf r})+\delta\hat n/\Omega$ so that
$\rho_0+\delta\hat\rho=\hat\rho+\delta\hat\sigma$. Note that $\hat\rho({\bf
r})$ also commutes with ${\bf \nabla}\hat\phi$. The
operator $\hat\psi_1({\bf r})$ appearing in Eq.(\ref{decompose2}) describes
short-wavelength components of $\hat\psi({\bf r})$  whose elimination
renormalizes the coefficients of the effective long-wavelength theory for
$\hat\phi(\bf r)$, $\delta\hat\rho(\bf r)$ \cite{Popov}, and which provide a
particle reservoir for the condensate even in the case $T=0$ to which we
confine
ourselves here. In the following we shall assume that the elimination of
$\hat\psi_1$ has already been performed. The
ansatz (\ref{decompose2}) is used in the Hamiltonian
(\ref{hamiltonian}) to determine the Heisenberg equation of motion for
$\delta\hat\rho$ and $\hat\phi$. We obtain formally (i.e. without paying
attention to the fact that operator products at equal points in space are only
well defined after chosing an apropriate regularization procedure; for our
present purposes this is sufficient, since we are only interested in the
quantized fields
$\delta\hat{\rho}, \delta\hat{\phi}$ either in their free-field regime, like in
the present section, or in the case where all modes except one are in their
vacuum states, as in the following section),
\begin{equation}
\delta\dot{\hat\rho}({\bf r},t) =
-(\hbar/m)\bbox\nabla\bbox\cdot \left[\left(\rho_0({\bf
r})+\delta\hat\rho({\bf
r},t)\right)^{1/2} \bbox\nabla\hat\phi({\bf r},t)\left(\rho_0({\bf
r})+\delta\hat\rho({\bf
r},t)\right)^{1/2}\right]\:,
\label {nl1} \end{equation}
\begin{eqnarray}{\nonumber}
\dot{\hat\phi}({\bf r},t) &=&
-\frac{1}{\hbar}(U_0(\rho_0({\bf r})+\delta\hat{\rho}({\bf r},t)) +
V+\delta V({\bf r},t)-\mu )\\
&& -\frac{\hbar}{4m}[\left(\rho_0({\bf
r})+\delta\hat\rho({\bf
r},t)\right)^{-1/2}({\bf \nabla} \hat{\phi})^2 \left(\rho_0({\bf
r})+\delta\hat\rho({\bf
r},t)\right)^{1/2}
\nonumber \\
&& + \left(\rho_0({\bf
r})+\delta\hat\rho({\bf
r},t)\right)^{-1/2}({\bf \nabla} \hat{\phi})^2 \left(\rho_0({\bf
r})+\delta\hat\rho({\bf
r},t)\right)^{-1/2}] \nonumber\\
&& + \frac{\hbar}{4m}\left(\rho_0({\bf
r})+\delta\hat\rho({\bf
r},t)\right)^{-1/2} (\nabla )^2 \left(\rho_0({\bf
r})+\delta\hat\rho({\bf
r},t)\right)^{1/2}\:,
\label {nl2} \end{eqnarray}

It follows from (\ref{nl1}) that
\begin{equation}
\delta\dot{\hat n}=\int d^3r\delta\dot{\hat\rho}=0\:,\label {n}
\end{equation}
i.e. there is no restoring
force on $\delta\hat{n}$, which therefore need not necessarily be small.
Therefore we combine this quantity with $\rho_0$ to extract from (\ref{nl2})
the equation which holds to zeroth-order in the small quantities ${\bf
\nabla}\hat\phi$ and $\delta\hat\sigma$
  \begin{equation}
 \hat\mu\hat\rho({\bf r}) =
-\frac{\hbar^2}{4m}\left(\nabla^2\hat\rho({\bf r}) -
\frac1{2}\hat\rho({\bf r})^{-1}\left(\bbox\nabla\hat\rho({\bf r})\right)^2
\right) +
V\hat\rho({\bf r}) + U_0\hat\rho({\bf r})^2 \:.\label {GP}
\end{equation}
This is one component of the Gross-Pitaevski equation and identifies
$\hat\rho({\bf
r})$ as the operator of the density of the condensate, as well as defining the
operator of the chemical potential $\hat\mu = \mu (N_0 +\delta\hat n)$ as a
function of the
number operator $ N_0+\delta\hat n= \int d^3r\hat\rho({\bf r})$ in the
condensate. Next we linearise with respect to the small quantities
$\delta\hat\sigma({\bf r},t)$ and $\bbox\nabla\hat\phi({\bf r})$
to obtain
\begin{equation} \delta\dot{\hat\sigma}({\bf r},t) = - \frac{\hbar}m
\bbox\nabla\bbox\cdot \left[ \hat\rho({\bf r})
 \bbox\nabla\hat\phi({\bf r},t)\right] \:, \label{sigmadot}
\end{equation}
\begin{equation} \dot{\hat\phi}({\bf r},t)=
-\frac1{\hbar}U_0\delta\hat\sigma({\bf r},t) +
\frac{\hbar}{4m}\hat\rho({\bf r})^{-1/2}\left(\nabla^2  -
\hat\rho({\bf r})^{-1/2}(\nabla^2\hat\rho({\bf r})^{1/2})\right)
\hat\rho({\bf r})^{-1/2}\delta\hat\sigma({\bf r},t)
-\frac1{\hbar}\delta V({\bf r},t) \:. \label{phidot}
\end{equation}

 Eliminating  $\bbox\nabla\hat\phi({\bf r})$ from
(\ref{sigmadot}) and (\ref{phidot}) we then get the wave equation for the
low-lying excitations,
\begin{eqnarray}{\nonumber}
\delta\ddot{\hat\sigma}({\bf r},t) &=&
\frac{1}m \bbox\nabla\bbox\cdot \left[
\hat\rho({\bf r})
\bbox\nabla\left(U_0\delta\hat\sigma({\bf r},t)
+\delta V({\bf
r},t)\right)\right]\\
&& -\frac{\hbar^2}{4m^2}\bbox\nabla\bbox\cdot\left[\hat\rho({\bf
r})\bbox\nabla\left[\hat\rho({\bf r})^{-1/2}\left(\nabla^2-\hat\rho({\bf
r})^{-1/2}(\nabla^2\hat\rho({\bf r})^{1/2})\right)\hat\rho({\bf r})^{-1/2}
\delta\hat\sigma({\bf r},t)\right ]\right ] \:.
\label{sigmaddot} \end{eqnarray}

Neglecting the gradient term of fourth-order
compared to those of second-order, and also neglecting $\delta\hat n$
compared to
$\rho_0$ and replacing $\delta\hat\sigma$ by the classical density fluctuation,
we recover Stringari's wave equation \cite{9} for the collective excitations of
a trapped Bose condensate.  We shall in the following also neglect
the higher-order gradient terms, but shall keep the number
fluctuation operator $\delta\hat n$, which is
implicit in $\hat \rho$, and examine its
consequences in Eq. (\ref{sigmaddot}). Since the short-wavelength
non-condensate components of the system act as a reservoir even at
$T=0$ these number fluctuations may reasonably be taken as Poissonian.
We can therefore assume that
the initial state of the system is given by a pure state or a mixture with a
Gaussian distribution over eigenstates of $\delta\hat n$, with zero mean
particle fluctuation and mean square deviation given by $\langle ( \Delta N_0
)^2 \rangle = N_0$, the average
number of particles in the condensate.  This corresponds to a Poissonian
distribution of the total particle number in the condensate $\delta\hat n +
N_0$ around its mean value $N_0\gg1$. We emphasize that due to the
entanglement of
the condensate with the other states of the system, which have been
eliminated,
the use of a mixture of $\delta \hat n$ eigenstates applies even to a single
realization of the experiment, and not only to an ensemble of realizations
corresponding to identically prepared experiments.
For each eigenvalue $\delta n$  of $\delta\hat n$ we have in principle to
determine the solution of Eq. (\ref{GP}) and of Eq. (\ref{sigmaddot})
corresponding to a given mode.  This gives the normal modes of the density
oscillations, after $\delta V({\bf r},t)$ has been switched off, and determines
their frequencies $\omega_{\nu}$ as a function
$\omega_{0\nu}(N_0+\delta\hat
n)$, where $\omega_{0\nu}(N_0)$ is the collective mode frequency
determined by the mean-field theory \cite{7,8,9,Singh}.
In the quantum ensemble defined by the initial
state only eigenvalues $\delta n$ of  $\delta\hat n$ which are very small
compared to $N_0$ are contained with appreciable weight.  Therefore it is
sufficient to expand $\omega_{\nu}$ to first-order in $\delta n$
\begin{equation}
\omega_\nu = \omega_{0\nu}(1+\gamma_\nu\delta n) \:, \label{16}
\end{equation}
with
\begin{equation} \gamma_\nu =
\frac{1}{\omega_{0\nu}}\frac{\partial{\omega_{0\nu}}}{\partial{N_0}} \:,
\label{freqshift} \end{equation}
a quantity determined by the solutions of the mean-field
theory \cite{7,8,9,Singh}.

We can now discuss the collapse and revivals of the collective excitations by
evaluating the quantum ensemble average of the
density oscillation $\langle\delta\hat\sigma_\nu(t)\rangle$ for a given mode,
assuming that it is coherently excited, e.g. by modulating the trap at the
required
frequency at times prior to $t=0$, while for $t>0$ the mode is left to evolve
freely.  We obtain at times $0<t<(\gamma_\nu\omega_{0\nu})^{-1}$
\begin{equation}
\langle\delta\hat{\sigma}_{\nu}(t)\rangle
= \exp(-\frac{1}{2}N_0\gamma_{\nu}^2\omega_{0\nu}^2t^2)
\left (\langle\delta\hat\sigma_{\nu}(0)
\rangle\cos (\omega_{0\nu}t)
+\frac{\langle\delta\dot{\hat\sigma}_{\nu}(0)\rangle}
{\omega_{0\nu}}\sin(\omega_{0\nu}t) \right )
\:, \label{collapseform}
\end{equation}
It can be checked that the corrections of order $(\delta n)^2$ to (\ref{16})
make a negligible contribution in (\ref{collapseform}).
Equation (\ref{collapseform}) shows that the excitation
decays by dephasing on a timescale
\begin{equation}
\tau_c=(\sqrt{N_0/2}|\gamma_\nu|\omega_{0\nu})^{-1}\:,
\label{tau} \end{equation}
in the form
of a Gaussian collapse. This collapse occurs even at a temperature $T=0$,
and even if the condensate never has a well-defined overall phase. This
collapse
is therefore different from (and turns out to be less effective than) the
collapse of the macroscopic wave function due to number-fluctuations
\cite{WriWalGar96}, which occurs if a well defined phase of the condensate
(with respect to some phase
standard, like another condensate) is prepared e.g. at $t=0$. On
timescales $t\ge (\gamma_\nu\omega_{0\nu})^{-1}$ the discreteness of the
spectrum of  $\delta\hat n$ manifests itself and we obtain revivals of
$\langle\delta\hat\rho_\nu(t)\rangle$ at the times
$t=n\pi/\gamma_\nu\omega_{0\nu}$.

To estimate the order of magnitude of the effect and to examine its
accessibility to observation we have to obtain an estimate of
$\gamma_{\nu}$ e.g. for
the observed $m=0$ modes in the experimentally realized condensates. We
can
achieve this goal by using some analytical results due to
Stringari \cite{9}. It was
shown in \cite{9} that for $N_0 \rightarrow \infty$ the mode frequencies
become
independent of $N_0$. It was also shown there by using sum-rule
arguments that
for finite $N_0$ the $\omega_{0\nu}$ for the low-lying states can be
represented
in the form $\omega_{0\nu}(N_0) =
\omega_{0\nu}(\infty)(1+c_{\nu}E_{kin}/E_{ho})$, where the $c_{\nu}$
are
parameters of order 1 which  depend on the trap geometry and the particular
mode, and $E_{kin}/ E_{ho}$ is the ratio of the kinetic energy and the
harmonic trap energy in the ground state. The $N_{0}$ dependence of this
ratio
in the limit of large $N_0$ of interest here can be estimated by using the
Thomas-Fermi approximation \cite {TF} as $E_{kin}/E_{ho}=b_{\nu}(N_0
a/\sqrt{\hbar/m\omega_0})^{-4/5}$ with $\omega_0$ the harmonic trap
frequency
and a proportionality factor $b_{\nu}$ of order 1 which is asymptotically
independent of $N_0$ but dependent on the trap geometry. For the
coefficient
$\gamma_{\nu}$ we obtain finally
\begin{equation} \gamma_{\nu}=-\frac {4}{5}
b_{\nu}c_{\nu}N_0^{-9/5}( a/d_0)^{-4/5}\:,\label {alpha}
\end{equation}
with $d_0 = \sqrt{\hbar/m\omega_0}$, giving a collapse time
\begin{equation}
\tau_c=(5\sqrt{2}/4b_{\nu}c_{\nu})(a/d_0)^{4/5}
\frac{N_0^{13/10}}{\omega_{0\nu}}\:,
\label {tauc} \end{equation}
which is larger than the collapse time for the macroscopic wave function obtained in \cite{WriWalGar96} roughly by a factor of $\mu/\omega_{0\nu}$.

For the two experiments in which decay times of collective excitations have
been
measured, the following parameters can be estimated:
In the experiment of
\cite{5}, the lifetime of the
$m=0$ mode (with a frequency of $\nu_{0\nu}=\omega_{0\nu}/2\pi=1.84
\nu_r$ where the radial trap frequency has the value $\nu_r =132 {\rm Hz}$)
for a condensate containing
$N_0=4500$ rubidium-87 atoms was measured to be $110 \pm 25 {\rm
ms}$. For this
case the following numbers apply:  $\omega_{0\nu}/2\pi=187Hz,
N_0=4500,
a=52 A , \omega_0=\omega_r (\omega_z/\omega_r)^{1/3},
a/\sqrt{\hbar/m\omega_0}=10^{-2}$. This gives a collapse time of
$\tau_c=850
{\rm ms}$, which is longer than the observed decay time, but this is to be
expected, as the latter was measured in a regime where finite temperature
effect are not negligible \cite{JinMatEns97}.

In the experiment of \cite{6}, the
lifetime of the $30 {\rm Hz}$ collective excitation of a condensate of
$N_0=5\times 10^6$ sodium atoms was measured to be $250 {\rm ms}$.
Due to the
much larger number of atoms used in this experiment the mode frequencies
$\omega_{0\nu}$ become insensitive to the dispersion in the number of
particles in the condensate \cite {9}, and the collapse due to the
mechanism under discussion here occurs only on  time-scales very long
compared to the observed damping time.

\section{Collapse due to nonlinearity of the collective modes}

In this section we neglect any uncertainty in the number of particles in
the condensate. However, we
no longer linearize the Heisenberg equations of motion for
$\delta\hat\rho$ and $\delta\hat\phi$. Instead we consider the case where
only a single mode of the collective excitations is appreciably excited
and ask for the nonlinear corrections to the dynamics of this mode. This nonlinearity will also give rise to a dephasing and collapse of the collective mode amplitude. This
is in the same spirit as the recent work of Kuklov et. al. \cite{Kuklov} and
Pitaevskii \cite{Pitaevskii}, but differs from the former in that we don't
treat a model but rather start from the full microscopic approach, and
is more general than the latter because
the analysis is not confined to a special symmetric collective mode.

The analysis starts with the nonlinear Heisenberg equations of motion
Eqs. (\ref{nl1}) and (\ref{nl2}) but with the
density gradient-terms neglected.  These can be derived from the
Hamiltonian \cite{Griffin}
\begin{equation}
H = \int d^3x\left\{ \frac{\hbar^2 \hat\rho}{2m} ({\bf \nabla}\hat\phi)^2 +
\frac{U_0}2 (\delta\hat\sigma)^2 +
\frac{\hbar^2}{2m}\delta\hat\sigma^{1/2}
({\bf\nabla}\hat\phi)^2\delta\hat\sigma^{1/2} \right\}_{\cal N} \:.
\end{equation}
Questions of operator ordering are not important at this stage.  In the end we
will choose normal ordering in the mode operators; this is the meaning of the
notation $\{\ldots\}_{\cal N}$.  We introduce a mode expansion using
the modes of the linearised hydrodynamics,
\begin{equation}
\delta\hat\sigma = i \sum_\nu{} '
\left(\sqrt\frac{\hbar\omega_{\nu}}{2U_0}F_\nu ({\bf r})\alpha_\nu(t) -{\rm
h.c.}\right)\:,
\end{equation}
\begin{equation}
{\bf \nabla}\hat\phi = \sum_\nu{} '
\left(\sqrt\frac{U_0}{2\hbar\omega_{\nu}}{\bf \nabla} F_\nu ({\bf r})
\alpha_\nu(t) +{\rm
h.c.}\right)\:,
\end{equation}
where $\sum'$ is the mode sum {\em not} including the spatially and
temporally
constant part; the constant part is already included in $\hat\rho
= \rho_0 + \delta\hat n/\Omega$.
Here $\omega_\nu$ is the frequency associated with
the linear mode function $F_\nu ({\bf r})$,
so that the linear hydrodynamics \cite{9} implies
$\alpha_\nu(t) = \alpha_\nu e^{-i\omega_\nu t}$.

The Hamiltonian now becomes
\begin{equation}
H = \sum_\nu{}'\hbar\omega_\nu\alpha^\dagger_\nu\alpha_\nu
+
\hbar\sum_{\nu\kappa\lambda}{}'\left\{
\frac13
C^{(1)}_{(\nu\kappa\lambda)}\alpha_\nu\alpha_\kappa\alpha_\lambda
+\frac12
C^{(2)}_{\nu(\kappa\lambda)}\alpha^\dagger_\nu\alpha_\kappa\alpha_
\lambda
+ {\rm h.c.} \right\} \:,
\end{equation}
with $C^{(1)}$ and $C^{(2)}$ symmetric in the
parenthesised indices and given by
\begin{equation}
C^{(1)}_{(\nu\kappa\lambda)} =
\frac{i\hbar}{4m}\sqrt{\frac{U_0}{2\hbar}}
\left( D^{(1)}_{\nu(\kappa\lambda)} + D^{(1)}_{\kappa(\nu\lambda)}
+ D^{(1)}_{\lambda(\nu\kappa)} \right)  ,
\end{equation}
with
\begin{equation}
D^{(1)}_{\nu(\kappa\lambda)} =
\sqrt{\frac{\omega_\nu}{\omega_\kappa\omega_\lambda}}
\int d^3x F_\nu \left(\nabla F_\kappa \cdot \nabla F_\lambda \right) \:,
\end{equation}
and
\begin{equation}
C^{(2)}_{\nu(\kappa\lambda)} =
\frac{i\hbar}{4m}\sqrt{\frac{U_0}{2\hbar}}
\left( 2D^{(2)}_{\nu(\kappa\lambda)}
+ D^{(3)}_{\kappa\nu\lambda} + D^{(3)}_{\kappa\lambda\nu}
+ D^{(3)}_{\lambda\nu\kappa} + D^{(3)}_{\lambda\kappa\nu} \right)  ,
\end{equation}
where
\begin{equation}
D^{(2)}_{\nu(\kappa\lambda)} =
\sqrt{\frac{\omega_\nu}{\omega_\kappa\omega_\lambda}}
\int d^3x F^*_\nu \left(\nabla F_\kappa \cdot \nabla F_\lambda \right) \:,
\qquad
D^{(3)}_{\nu\kappa\lambda} =
\sqrt{\frac{\omega_\nu}{\omega_\kappa\omega_\lambda}}
\int d^3x F^*_\nu \left(\nabla F^*_\kappa \cdot \nabla F_\lambda \right) \:.
\end{equation}
The resulting equations of motion in the interaction representation are
\begin{equation}\label{interactionrep}
\dot\alpha_\nu = -i \sum_{\kappa\lambda}{}'\left\{
C^{(1)*}_{(\nu\kappa\lambda)}e^{i(\omega_\nu+\omega_\kappa+\omega_\lambda)t}
\alpha^\dagger_\kappa\alpha^\dagger_\lambda
+ \frac12 C^{(2)}_{\nu(\kappa\lambda)}
e^{i(\omega_\nu-\omega_\kappa-
\omega_\lambda)t}\alpha_\kappa\alpha_\lambda
+ C^{(2)*}_{\kappa(\nu\lambda)}e^{i(\omega_\nu-
\omega_\kappa+\omega_\lambda)t}
\alpha^\dagger_\lambda\alpha_\kappa
\right\} \:.
\end{equation}

We now assume that only a single mode $\mu$ is externally excited;
all other modes are only excited via their coupling to the mode $\mu$.
This implies that $\alpha_\mu$ can be considered ``large'' compared to all
other mode operators.  Using Eq. (\ref{interactionrep}) in this
approximation, we have for  $\nu=\mu$
\begin{equation}\label{mueqn}
\dot\alpha_\mu \simeq -i \sum_{\kappa}{}'\left\{
2C^{(1)*}_{(\mu\mu\kappa)}e^{i(2\omega_\mu+\omega_\kappa)t}
\alpha^\dagger_\mu\alpha^\dagger_\kappa
+ C^{(2)}_{\mu(\mu\kappa)}
e^{-i\omega_\kappa t}\alpha_\mu\alpha_\kappa
+ C^{(2)*}_{\mu(\mu\kappa)}
e^{i\omega_\kappa t}\alpha^\dagger_\kappa\alpha_\mu
+ C^{(2)*}_{\kappa(\mu\mu)}
e^{i(2\omega_\mu-\omega_\kappa)t}\alpha^\dagger_\mu\alpha_\kappa
\right\} \:,
\end{equation}
and for  $\nu=\kappa\ne\mu$
\begin{equation}\label{kappaeqn}
\dot\alpha_\kappa \simeq -i \left\{
C^{(1)*}_{(\kappa\mu\mu)}e^{i(\omega_\kappa+2\omega_\mu)t}
(\alpha^\dagger_\mu)^2
+ \frac12 C^{(2)}_{\kappa(\mu\mu)}
e^{i(\omega_\kappa-2\omega_\mu)t}(\alpha_\mu)^2
+ C^{(2)*}_{\mu(\kappa\mu)}
e^{i\omega_\kappa t}\alpha^\dagger_\mu\alpha_\mu
\right\} \:.
\end{equation}
In general, the oscillatory terms in this will be changing much
more rapidly than $\alpha_\mu$ in the interaction picture.  The only
exception to this is when we have a
second harmonic resonance, $\omega_\kappa = 2\omega_\mu$; this case
needs
separate
treatment, and is discussed below.  There is no contribution from the
apparent
resonance when $\omega_\kappa=0$, since this mode is explicitly excluded
from
the sum; in fact, the coefficient $C^{(2)*}_{\mu(0\mu)}$ would be zero
anyway.
Thus we may solve Eq. (\ref{kappaeqn}) for $\alpha_\kappa$ treating
$\alpha_\mu$
as approximately constant, to find
\begin{equation}\label{kappasoln}
\alpha_\kappa \simeq - \left\{
\frac{C^{(1)*}_{(\kappa\mu\mu)}}{\omega_\kappa+2\omega_\mu}
e^{i(\omega_\kappa+2\omega_\mu)t}(\alpha^\dagger_\mu)^2
+ \frac12 \frac{C^{(2)}_{\kappa(\mu\mu)}}{\omega_\kappa-2\omega_\mu}
e^{i(\omega_\kappa-2\omega_\mu)t}(\alpha_\mu)^2
+ \frac{C^{(2)*}_{\mu(\kappa\mu)}}{\omega_\kappa}
e^{i\omega_\kappa t}\alpha^\dagger_\mu\alpha_\mu
\right\} \:.
\end{equation}
We now subsitute this back into Eq. (\ref{mueqn}) for $\alpha_\mu$, and
keep
only the nonoscillatory terms, to obtain
\begin{equation}
\dot\alpha_\mu = i\kappa \alpha^\dagger_\mu \alpha_\mu \alpha_\mu \:,
\end{equation}
where
\begin{equation}\label{kappa}
\kappa = 2\sum_{\kappa}{}'\left\{
\frac{{|C^{(1)}_{(\kappa\mu\mu)}|}^2}{\omega_\kappa+2\omega_\mu}
+ \frac{{|C^{(2)}_{\mu(\kappa\mu)}|}^2}{\omega_\kappa}
+ \frac14 \frac{{|C^{(2)}_{\kappa(\mu\mu)}|}^2}{\omega_\kappa-
2\omega_\mu}
\right\} \:.
\end{equation}
The form of the matrix elements (22)-(24) implies selection rules for the modes $\kappa$ contributing to the sum in Eq.(31). For axially symmetric trap potentials with inversion symmetry only modes contribute with positive parity and azimuthal quantum numbers $m_\kappa$ satisfying $m_\kappa=2m_\mu$ or $m_\kappa=0$ (via $C^{(2)}_{.(..)} \ne 0$) or $m_\kappa=-2m_\mu$ ( via $C^{(1)}_{(\kappa\mu\mu)} \ne 0$).

The effective self-coupling coefficient $\kappa$ experimentally manifests
itself by an energy dependence of the observed mode frequency according to
$\omega_\mu (E_\mu ) = \omega_\mu -(\kappa / \hbar \omega_\mu )E_\mu$.
Having derived (\ref {kappa}), the collapse and revival of the mode $\mu$
follow immediately: If at $t=0$ the mode is excited e.g. in a coherent state
with amplitude $A$ then for $t > 0$ its amplitude changes according to
\begin{equation}
\langle A|\alpha_\mu|A\rangle = A\exp\left[-|A|^2(1-\cos(\kappa t))\right
]\left
[\cos(|A|^2\sin(\kappa t))-i\sin(|A|^2\sin(\kappa t)) \right] \:.
\end{equation}
which for times $|\kappa| t<<1$ collapses with the new collapse time $\tau_c=(|A\kappa|)^{-1}$
according to $\langle A|\alpha_\mu|A\rangle \approx A\exp(-
\frac{1}{2}(t/t_c)^2$ but is revived at revival times $t_r=2n\pi\kappa ^{-1}$
for integers $n\ge 1$.

We can estimate the order of magnitude of the effective coupling constant
$\kappa$, in particular its scaling with the system parameters and the number
of atoms in the condensate, by using the Thomas-Fermi approximation
\begin{equation}
\kappa \sim \frac{|C|^2}{\omega_0} \sim
\frac{\hbar U_0}{m^2\omega_0} \frac1{\omega_0} \left(\frac1{r_{\rm
TF}}\right)^7
\sim \omega_0 N_0^{-7/5} \left( \frac{d_0}a \right)^{2/5} \:.
\end{equation}
which gives for the collapse time the estimate
\begin{equation}
\tau_c \sim (A\omega_0)^{-1} N_0^{7/5} \left( \frac{a}d_0 \right)^{2/5} \:.
\end{equation}
In this result, derived for an initially excited coherent state, we may replace the coherent amplitude $A$ by the variance of the excited quantum number to generalize it for an arbitrary initially excited state. The scaling of $\kappa$ with the system parameters is consistent with
Pitaevskii \cite{Pitaevskii} who derived the coupling coefficient for  a
special
mode. The scaling of the coefficient with an inverse power of $N_0$ tends
to make it small in most cases. Putting in the numbers for the two experiments,
assuming the excited quantum number $N_e$ to be $1\%$ of the total particle number
we obtain
\begin{eqnarray}
Rb: N_e &=& 45,\qquad  A = \sqrt{45},  \qquad\tau _c = 2.2 s \nonumber\\
Na:  N_e &=& 5\cdot 10^4,\  A = 10^2 \sqrt{5},\quad  \tau _c
= 1.47 \cdot 10^3 s .
\end{eqnarray}%
which is longer than the observed decay

Relatively larger values are obtained for $\kappa$ (and shorter ones for $\tau_c$) if there is a
collective mode
near the second harmonic of the externally excited mode.
Second-harmonic generation has recently been seen in numerical simulations of the Gross-Pitaevskii equation \cite{Burnett}.

We shall briefly discuss this point for the case of an anisotropic axially symmetric trap whose collective mode frequencies and eigenfunctions are known as a function of the anisotropy parameter $\beta=\omega_z/\omega_0$ in the Thomas-Fermi and hydrodynamic limit (\cite{9},\cite{Fliesser}), where $\omega_z, \omega_0$ are the axial and radial trap frequencies, respectively. The modes are in this case labelled by three quantum numbers $(n, j, m)$ (see \cite{Fliesser}). We are interested in the case where for two different sets of their values we have second harmonic resonance such that
$\omega_{\nu}=2\omega_{\mu}$, {\it and} where the relevant matrix element
$C^{(2)}_{\nu (\mu \mu )}$ with $\nu=(\bar n, \bar j, \bar m), \mu=(n, j, m)$
does not vanish. The latter condition implies certain selection rules, which for the case at hand read: i) The mode function $F_\nu ({\bf r})$ of the second harmonic mode in cylindrical coordinates $\rho,z,\varphi$ must be even in $z$, and ii) at least one of the two conditions $\bar m = 2m$ (for $D^{(2)}_{\nu (\mu \mu)} \ne 0, D^{(3)}_{\mu \mu \nu} \ne 0$) or $\bar m =0$(for $D{(3)}_{\mu \nu \mu} \ne 0$) must be satisfied.

As an example we consider the mode $\mu = (n=2, j=0, m=2)$ which has $\omega_{\mu} = \sqrt{2} \omega_0 , F_{\mu}({\bf r}) = N_{\mu}\rho^2 \exp (2i\varphi)$ where $N_{\mu}$ is a normalization factor, and we shall fix the phases of the mode functions by chosing these normalization factors always positive. Then the two modes $\nu1=(\bar n=2, \bar j=1, \bar m =0)$ with $\omega_{\nu1}^2 =(3\beta^2/2+2+\sqrt{(2-3\beta^2/2)^2+2\beta^2})\omega_0^2, F_{\nu1}({\bf r})=N_{\nu1}(-4\rho^2/3-16z^2/3+1)$ and $\nu_2=(\bar n=2, \bar j=0, \bar m=4)$ with $\omega_{\nu2}^2 =(3\beta^2/2+10-\sqrt{(6-3\beta^2/2)^2+10\beta^2})\omega_0^2, F_{\nu2}({\bf r})=N_{\nu2}(-16z^2+2\rho^2-1)\rho^4\exp (4i\varphi)$ both satisfy the selection rules. For the special value
$\beta^2=(\omega_z/\omega_0)^2=16/7$ it turns out that both modes $\nu1, \nu2$ are degenerate and in second harmonic resonance with the mode at frequency $\omega_{\mu}=\sqrt{2}\omega_0$, i.e. $\omega_{\nu1}=\omega_{\nu2}=2\sqrt{2}\omega_0$. (A resonance for this value of $\beta^2$ has recently also been noted in \cite{Min}). The relevant matrix elements in this case turn out to be $C^{(2)}_{\nu1 (\mu \mu)} = 0.562 N_0^{-7/10}(d_0/a)^{1/5}\omega_0$ and $C^{(2)}_{\nu2 (\mu \mu)} =  0.332 N_0^{-7/10}(d_0/a)^{1/5}\omega_0$
For such a case of exact second harmonic resonance, the previous treatment
breaks
down. The differential equations for the two, or, in the present case even three, resonantly coupled modes must
instead be
considered together.
Neglecting couplings to all other modes, we have for $\omega_{\nu i} =
2\omega_\mu$
\begin{equation}
\dot\alpha_\mu = -i\sum_i C^{(2)*}_{\nu i(\mu\mu)} \alpha^\dagger_\mu \alpha_{\nu i} \:,
\end{equation}
\begin{equation}
\dot\alpha_{\nu i} = -\frac i2 C^{(2)}_{\nu i(\mu\mu)} \alpha_\mu^2 \:,
\end{equation}
which are the coupled mode equations for second harmonic or subharmonic
generation.  It can be seen that only the linear combination $\alpha_{\nu}=const.\sum_i C^{(2)*}_{\nu i (\mu \mu)}\alpha_{\nu i}$ of second harmonic modes couples to the fundamental mode $\mu$, while the linear combination $\bar \alpha_{\nu}=const.[C^{(2)}_{\nu_2 (\mu \mu)}\alpha_{\nu1}-C^{(2)}_{\nu1 (\mu \mu)}\alpha_{\nu2}]$ is not generated. Choosing the constants ($const=|C^{(2)}_{\nu (\mu\mu)}|^{-1}$) to normalize these linear combinations, Eqs.(36),(37) reduce to
\begin{equation}
\dot\alpha_\mu = -i\|C^{(2)}_{\nu(\mu \mu)}| \alpha^\dagger_\mu \alpha_\nu
\:,
\end{equation}
\begin{equation}
\dot\alpha_\nu = -\frac i2 |C^{(2)}_{\nu (\mu\mu)}| \alpha_\mu^2 \:,
\end{equation}
where we defined $|C^{(2)}_{\nu (\mu\mu)}|=[|C^{(2)}_{\nu1(\mu\mu)}|^2+|C^{(2)}_{\nu2(\mu\mu)}|^2]^{1/2}$, which takes the value $|C^{(2)}_{\nu (\mu\mu)}|=0.653 N_0^{-7/10}(d_0/a)^{1/5}\omega_0$ in our example.
These equations have well-known exact solutions \cite{Bloembergen} in
terms of
Jacobi elliptic
functions, that oscillate with a frequency
\begin{equation}
\Omega_0 =
|C^{(2)}_{\nu(\mu\mu)}|\sqrt{\langle\alpha^\dagger_\mu\alpha_\mu\rangle}
\:.
\end{equation}
An order-of-magnitude estimate of this is
\begin{equation}
\Omega_0 \sim  \sqrt{\frac{n_\mu \hbar U_0}{m^2\omega_0 r_{\rm TF}^{7}
}}
\sim \sqrt{n_\mu} \omega_0 N_0^{-7/10} \left( \frac{d_0}a
\right)^{1/5} \:,
\end{equation}
which is much larger than the nonresonant rate $\kappa \sqrt{n_\mu}$. The
initial transfer of energy from the fundamental mode to the second harmonic
may be viewed as a collapse with a collapse time $\tau_c\sim\Omega_0$.

The resonance phenomenon found here might help to explain why in the
experiments some modes show
amplitude dependence of the frequency, while others do not.

\section{Conclusions}
We have investigated two different quantum mechanisms which may lead to
a dephasing of collective modes in trapped Bose-Einstein condensates even at
zero temperature. Both rely on the possibility that the system may be in a
linear superposition of states each of which has a slightly different
frequency
for the collective mode. In the first case investigated here the linear
superposition combines states with different numbers of atoms in the
condensate. This may occur since the non-condensate atoms, which are always
present due to many-body interactions \cite{Bog47},
act as a particle reservoir for the
condensate. This case was earlier shown to give rise to collapses and revivals
of the macroscopic wave-function \cite {WriWalGar96}, once the latter is
prepared at a given time with a well-defined (relative) phase, e.g. by a
measurement. Here we found that a dephasing effect occurs also for the
collective mode but on a different time scale which is typically much
longer for large condensates.  In principle, the collapse of the macroscopic wave-function, if it was initially prepared with a well-defined phase, and of the collective mode amplitude may occur independently and simultaneously, but their simultaneous observation would of course be much more difficult than the observation of each collapse individually. While the dimensionless collapse time of
the macroscopic wave function for large $N_0$ scales like
\begin{equation}
\omega_0 \tau_c \sim (d_0/a)^{2/5} N_0^{3/5}/\langle \Delta N_0 \rangle
\,\mbox{\rm (macroscopic wave function)}
\end{equation}
the collapse time of the collective mode based on this mechanism scales as
\begin{equation}
\omega_0 \tau_c \sim (d_0/a)^{-4/5} N_0^{9/5}/\langle \Delta N_0 \rangle
\,\mbox{ \rm (collective mode due to atom-number uncertainty).}
\end{equation}
which corresponds to Eq.(17), which was written for $\langle\Delta N_0 \rangle
=N_0^{1/2}$.

In the second case investigated by us the linear superposition is one of
different quantum number states of the excited collective mode, which will
typically be in a coherent state immediately after its excitation from its
vacuum state. We have calculated the effective nonlinearity of the excited
mode due to its coupling to the other modes. This nonlinearity gives rise to a
dispersion of the collective mode frequency within the linear superposition of
number states. The collapse time obtained from this mechanism we find (in Eq.(34)) to
scale like
\begin{equation}
\omega_0 \tau_c \sim (d_0/a)^{-2/5} N_0^{7/5}/\langle \Delta n \rangle
\,\mbox{ \rm (collective mode due to nonlinearity)}
\end{equation}
For an initially excited coherent state the variance of the quantum number is
related to the average mode-energy $E_\mu$ by $\langle\Delta n \rangle =
\sqrt{E_\mu / \hbar \omega_\mu}$. This scaling is consistent with a result
obtained by Pitaevskii \cite{Pitaevskii}. We have also found that the
nonlinearity may be strongly enhanced if there is a mode at or close to a
resonance with the second harmonic of the excited mode. In the case of such
a resonance the transfer of energy from the fundamental mode to the second
harmonic for short times looks similar to a collapse of the fundamental mode,
which, according to Eq.(41), occurs on a time-scale
scaling like
\begin{equation}
\omega_0 \tau_c \sim (d_0/a)^{-1/5} N_0^{7/10}/\sqrt{E_\mu / \hbar
\omega_\mu}
\,\mbox{ \rm (collective mode due to second-harmonic)}.
\end{equation}
Due to the scaling of the collapse times with positive powers of the atom
number the collapse by one of the mechanisms we have examined could best
be observed in comparatively small condensates. At presently achieved
temperatures the observed damping of the collective mode occurs at a higher
rate than the collapse rates obtained here. However the observed damping
rates were found to be strongly temperature dependent, decreasing rapidly
with decreasing temperatures \cite{JinMatEns97}.
Therefore the collapse we have
investigated here is still masked in present experiments by finite temperature
effects, but is predicted to reveal itself when experiments are pushed to
smaller temperatures.

\vspace{0.25cm}
\noindent
R.G. acknowledges support from the Deutsche Forschungsgemeinschaft
through
SFB 237;
E.M.W. is partially supported by the Joint Services Optical Program;
D.F.W. and M.J.C. acknowledge support from the Marsden Fund of the
Royal
Society
of New Zealand, the University of Auckland Research Committee and the
New
Zealand Lotteries' Grants Board; and D.F.W. also from the Office of Naval
Research.

\end{document}